\documentclass[12pt]{iopart}
\usepackage{iopams}
\usepackage{graphicx}
\expandafter\let\csname equation*\endcsname\relax
\expandafter\let\csname endequation*\endcsname\relax
\usepackage{amsmath}
\usepackage{amssymb}
\usepackage{cite}

\begin{document}

\title[The mean and variance of the distribution of shortest path lengths]
{The mean and variance of the distribution of shortest path lengths of random regular graphs  
}

\author{Ido Tishby$^1$, Ofer Biham$^1$, Reimer K\"uhn$^2$ and Eytan Katzav$^1$}
\address{$^1$Racah Institute of Physics, The Hebrew University, Jerusalem 9190401, Israel}
\address{$^2$Department of Mathematics, King's College London, Strand, London WC2R 2LS, UK}
\eads{\mailto{ido.tishby@mail.huji.ac.il}, 
\mailto{biham@phys.huji.ac.il}, 
\mailto{reimer.kuehn@kcl.ac.uk},
\mailto{eytan.katzav@mail.huji.ac.il}}

\begin{abstract}

The distribution of shortest path lengths (DSPL) of random networks provides
useful information on their large scale structure. In the special case
of random regular graphs (RRGs), which consist of $N$ nodes of degree $c \ge 3$,
the DSPL, denoted by $P(L=\ell)$, follows a discrete Gompertz distribution.
Using the discrete Laplace transform
we derive a closed-form expression for the moment generating function
of the DSPL of RRGs.
From the moment generating function we obtain closed-form expressions for 
the mean and variance of the DSPL.
More specifically, we find that the mean distance between pairs 
of distinct nodes is given by
$\langle L \rangle =   
\frac{\ln N}{\ln (c-1)} 
+ \frac{1}{2}
- \frac{ \ln c - \ln (c-2) +\gamma}{\ln (c-1)}
+ \mathcal{O} \left(  \frac{\ln N}{N}  \right)$,
where $\gamma$ is the Euler-Mascheroni constant. 
While the leading term is known, this result includes a novel
correction term, which yields very good agreement with 
the results obtained from direct numerical evaluation of $\langle L \rangle$
via the tail-sum formula and with the results obtained from computer simulations.
However, it does not account for an oscillatory behavior of $\langle L \rangle$
as a function of $c$ or $N$. 
These oscillations are negligible in sparse networks but 
detectable in dense networks.
We also derive an expression for the variance ${\rm Var}(L)$ of the DSPL,
which captures the overall dependence of the variance on $c$
but does not account for the oscillations.
The oscillations are due to the discrete nature of the shell structure around a random node.
They reflect the profile of the filling of new shells as $N$ is increased.
The results for the mean and variance are compared to the corresponding results 
obtained in other types of random networks.
The relation between the mean distance and the diameter is discussed.

\end{abstract}

%\pacs{05.40.Fb, 64.60.aq, 89.75.Da}


%\submitto{\jpa (\today)}
%\date
\maketitle

\section{Introduction}

Random networks (or graphs) consist of a set of $N$ nodes that
are connected by edges in a way that is determined by some
random process.
They provide a useful conceptual framework 
for the study of a large variety of systems and processes
in science, technology and society
\cite{Dorogovtsev2003book,Havlin2010,Newman2010,Estrada2011,Barrat2012}.
The local structure of a random network
can be characterized by the degree distribution $P(k)$
and its moments $\langle K^n \rangle$, $n \ge 1$.
In particular, the mean degree $\langle K \rangle$ provides
the expected number of neighbors of a random node,
while the variance ${\rm Var}(K)=\langle K^2 \rangle - \langle K \rangle^2$
accounts for the width of the degree distribution.

The large scale structure of a random network is captured 
by the distribution of shortest path lengths (DSPL),
or the distance distribution, between pairs of distinct nodes.
Properties of the DSPL, which is denoted by $P(L=\ell)$, 
have been studied in random networks with different degree distributions
\cite{Newman2001,Dorogovtsev2003DSPL,Hofstad2005,Blondel2007,Hofstad2007,Hofstad2008,Esker2008,Shao2008,Shao2009,Goldental2015,Goldental2017,Katzav2015,Nitzan2016,Melnik2016,Bonneau2017,Steinbock2017,Steinbock2019,Katzav2018,Tishby2018,Asher2020,Jackson2021}.
It was shown that in random networks whose degree distribution has a finite variance,
the mean distance between pairs of distinct nodes scales like
$\langle L \rangle \sim \ln N$
\cite{Newman2001,Dorogovtsev2003DSPL,Esker2008}.
This implies that random networks are small-world networks
\cite{Travers1969,Watts1998,Chung2002,Chung2003,Fronczak2004}.
Moreover, it was shown that scale-free networks,
which exhibit a power-law degree distribution of the 
form $P(k) \sim k^{- \gamma}$, may be
ultrasmall depending on the value of the exponent $\gamma$.
In particular, for $2 < \gamma < 3$, where the variance of $P(k)$ diverges in the infinite system limit,
the mean distance scales like
$\langle L \rangle \sim \ln \ln N$
\cite{Cohen2003,Dorogovtsev2003DSPL,Hofstad2007}.
The variance ${\rm Var}(L)$ of the DSPL
was also studied. It was shown that the DSPL of random networks
is typically a narrow distribution, whose width does not grow as the
network size is increased
\cite{Dorogovtsev2003DSPL}.
 
In the study of the DSPL it is convenient to use the
tail distribution $P(L>\ell)$, which is the probability that
the distance between a random pair of distinct nodes $i$
and $j$ is larger than $\ell$.
In networks that consist of more than one connected component,
the distance between nodes that reside on different network components
is $\ell=\infty$.
In this case the DSPL is restricted to pairs of nodes that reside on the same
connected component
\cite{Katzav2015,Nitzan2016,Katzav2018,Tishby2018}.
In the analysis below we focus on networks that consist of a single
connected component.
In this case, the mean distance can be obtained from the tail-sum formula
\cite{Pitman1993}

\begin{equation}
\langle L \rangle = \sum_{\ell=0}^{\infty} P(L>\ell).
\end{equation}

In configuration model networks the degree of each node is 
drawn independently from a given degree distribution $P(k)$ and
the connections are random and uncorrelated
\cite{Molloy1995,Molloy1998,Newman2001}.
The configuration model generates maximum entropy ensembles 
in which the degree distribution $P(k)$ is fixed
\cite{Newman2001,Fronczak2004,Molloy1995,Molloy1998}.
Therefore, the configuration model provides a general and highly powerful
platform for the analysis of statistical properties of networks.
In configuration model networks, the DSPL is completely determined
by the degree distribution $P(k)$ and the network size $N$.
In particular, in the large network limit the mean distance  can be approximated by
\cite{Newman2001,Chung2002,Chung2003,Fronczak2004}

\begin{equation}
\langle L \rangle
\simeq \frac{ \ln N }{ \ln \mu },
\label{eq:<L>conf}
\end{equation}

\noindent
where

\begin{equation}
\mu =
\frac{ \langle K^2 \rangle - \langle K \rangle }{\langle K \rangle}
\end{equation}

\noindent
is the mean of the excess degree distribution of nodes selected via a random edge.
The excess degree distribution, given by
\cite{Newman2003}

\begin{equation}
P_{\rm excess}(k) = \frac{(k+1)}{\langle K \rangle} P(k+1),
\end{equation}

\noindent
is obtained by selecting random edges and choosing randomly one of the two end-nodes 
of the selected edge. The excess degree of such end-node is obtained by extracting
the edge that led to that node, reducing its degree by $1$.

The random regular graph (RRG) is a special case of a configuration 
model network, in which the degree distribution is a degenerate
distribution of the form 
$P(k)=\delta_{k,c}$, namely
all the nodes are of the same degree $c$,
where $c \ge 3$.
In the special case of an RRG,
the second moment of the degree distribution satisfies
$\langle K^2 \rangle = c^2$ and
the excess degree is $\mu = c - 1$.
As a result, Eq. (\ref{eq:<L>conf}) is reduced to

\begin{equation}
\langle L \rangle \simeq \frac{\ln N}{\ln(c-1)}.
\label{eq:Leading}
\end{equation}
 
Recently, the DSPL of the
Erd{\H o}s-R\'enyi (ER) network
\cite{Erdos1959,Erdos1960,Erdos1961}
and other configuration model networks 
was calculated using an approach 
called the random path approach (RPA), which is
based on recursion equations
\cite{Katzav2015,Nitzan2016,Bonneau2017}.
In general, the recursion equations are iterated step by step and the
resulting distribution is evaluated numerically.  
The DSPL obtained from the recursion equations was found to 
be in very good agreement with the results obtained from computer simulations
for configuration model networks with a broad range of degree distributions
\cite{Nitzan2016}. 
In the special case of random regular graphs (RRGs) 
of degree $c \ge 3$,
the recursion equations yield a
closed-form expression for the DSPL,
whose tail distribution takes the form
\cite{Nitzan2016}  

\begin{equation}
P(L> \ell) = \exp \left\{ -   \frac{c}{(c-2)N}   \Big[ (c-1)^{\ell} - 1 \Big] \right\}.
\label{eq:taildist0}
\end{equation}

\noindent
The distribution shown in Eq. (\ref{eq:taildist0}) is a discrete form of the
Gompertz distribution
\cite{Gompertz1825,Shklovskii2005}.
Eq. (\ref{eq:taildist0}) coincides with an earlier calculation of the DSPL of RRGs 
\cite{Hofstad2005}.
The existence of a closed-form expression for $P(L>\ell)$ opens the way to
the derivation of compact formulae for the mean distance $\langle L \rangle$
and the variance ${\rm Var}(L)$ of the DSPL of RRGs. 
The mean distance $\langle L \rangle$ provides the typical length scale in the
network, as well as its scaling with respect to the network size $N$ and the degree $c$.
The variance ${\rm Var}(L)$ provides the width of the peak 
in the probability mass function $P(L=\ell)$.

In this paper we derive a closed-form expression for the moment generating function
of the DSPL of RRGs.
The moment generating function is expressed in terms of a discrete Laplace transform.
The discrete sum is then evaluated using the Euler-Maclaurin formula
that applies for sufficiently large networks.
Using the moment generating function we obtain closed-form expressions for 
the mean distance $\langle L \rangle$ and for the second moment $\langle L^2 \rangle$ of the DSPL.
Using these results we also obtain,
for the first time, a closed form expression for the variance
${\rm Var}(L)$.
In previous studies
\cite{Newman2001,Chung2002,Hofstad2005,Shimizu2020}
it was found that in the large $N$ limit the mean distance $\langle L \rangle$ can be
approximated by Eq. (\ref{eq:Leading}).
On top of this result, we obtain a novel correction term,
which is of order $1$, namely independent of $N$.
Since the mean distance $\langle L \rangle$ is logarithmic in $N$,
such correction of order $1$ is often not negligible.
Taking this correction into account yields a very good agreement with  
the results obtained from computer simulations.
However, it does not account for an apparent oscillatory behavior of $\langle L \rangle$
as a function of $c$ and as a function of $N$.
These oscillations are negligible in the sparse-network limit but detectable in the dense-network limit.
The closed-form expression obtained for ${\rm Var}(L)$ captures the
overall dependence of the variance on the degree $c$.
However, it does not account for the oscillatory behavior of ${\rm Var}(L)$ as a
function of $c$, which becomes significant in the dense-network limit.
The oscillations of $\langle L \rangle$ and ${\rm Var}(L)$ as a function of $c$ and $N$ are analyzed and discussed.
In particular, it is shown that the oscillations become regular when plotted as a function of 
$\ln N/\ln(c-1)$, when $c$ is kept fixed and $N$ is varied.
The results for the mean and variance of the DSPL are compared to the corresponding results
obtained in other types of random networks.
The relation to the diameter is also discussed.

The paper is organized as follows.
In Sec. 2 we describe the random regular graph.
In Sec. 3 we review the calculation of the distribution of shortest path lengths.
In Sec. 4 we derive a closed-form expression for the moment  
generating function of the DSPL.
In Sec. 5 we calculate the mean distance $\langle L \rangle$.
In Sec. 6 we calculate the variance of the DSPL.
The results are discussed in Sec. 7 and summarized in Sec. 8.

\section{The random regular graph}

The RRG is a special case of a configuration 
model network, in which the degree distribution is a degenerate
distribution of the form 
$P(k)=\delta_{k,c}$, namely
all the nodes are of the same degree $c$.
Here we focus on the case of $3 \le c \le N-1$,
in which for a sufficiently large value of $N$ the RRG consists of a single connected component
\cite{Molloy1995,Molloy1998,Bollobas2001}.
RRGs of any finite size exhibit a local tree-like structure,
while at larger scales there is a broad spectrum of cycle lengths.
In that sense RRGs differ from Cayley trees, which maintain their
tree structure by reducing the most peripheral nodes to leaf nodes of degree $1$.
They also differ from Bethe lattices which exhibit a tree structure of an infinite size.

The neighborhood of a given node $i$ can be described by a shell structure, in which
the first shell consists of the $c$ neighbors of $i$ and the second shell consists of the
neighbors of the nodes of the first shell, which are at distance $\ell=2$ from $i$.
In general, the $\ell$th shell around $i$ consists of all the nodes that are at a distance
$\ell$ from $i$
\cite{Shao2008,Shao2009}.
In the infinite network limit, the $\ell$th shell around each node consists of 
$n(\ell) = c (c-1)^{\ell-1}$ nodes.

A convenient way to construct an RRG
of size $N$ and degree $c$
($Nc$ must be an even number)
is to prepare the $N$ nodes such that each node is 
connected to $c$ half edges or stubs
\cite{Newman2010,Coolen2017}.
At each step of the construction, one connects a pair of random stubs that 
belong to two different nodes $i$ and $j$ 
that are not already connected,
forming an edge between them.
This procedure is repeated until all the stubs are exhausted.
The process may get stuck before completion in case that
all the remaining stubs belong to the
same node or to pairs of nodes that are already connected.
In such case one needs to perform some random reconnections
in order to complete the construction.

\section{The distribution of shortest path lengths}

Consider an RRG consisting of $N$ nodes of degree $c$.
The distance $\ell_{ij}$ between a pair of nodes $i$ and $j$
is given by the length of the shortest path between $i$ and $j$.
The tail distribution of shortest path lengths between pairs of random
nodes is denoted by $P(L > \ell)$.
In computer simulations this distribution is obtained by generating
a large number of network instances from
an ensemble of RRGs of a given size $N$ and degree $c$.
In each network instance one repeatedly selects pairs of random nodes, 
finds the shortest paths between them and measures their lengths.
These results yield the distribution $P(L>\ell)$.

Another useful way to sample nodes is via a random edge. 
In this case one selects a random edge $e$ and picks randomly
one of the end-nodes $\tilde i$ of the selected edge.
The edge $e$ is then deleted, giving rise to a reduced network
that includes all the nodes and edges, apart from $e$,
also called the cavity graph
\cite{Mezard2003}.
The tail distribution of shortest path lengths between pairs of distinct nodes 
consisting of a node $\tilde i$, 
selected via a random edge $e$, 
and a random node $j$,  
on the reduced network from which $e$ is removed, 
is denoted by $\widetilde P(L > \ell)$.

Apart from the tail distributions 
$P(L>\ell)$ and $\widetilde P(L>\ell)$,
it is also useful to consider various conditional distributions.
Consider a pair of random nodes $i$ and $j$, such that the distance
between them is known to be larger than $\ell-1$.
The conditional probability $P(L>\ell | L>\ell-1)$ provides the
probability that the length of the shortest path between $i$ and $j$
is larger than $\ell$, given that it is larger than $\ell-1$.
Similarly, for a pair of nodes consisting of a node $\tilde i$,
which is selected via a random edge $e$, 
and a random node $j$, 
the conditional probability $\widetilde P(L>\ell | L>\ell-1)$
is the probability that the distance between $\tilde i$ and $j$
is larger than $\ell$, given that it is larger than $\ell-1$,
in the reduced network from which $e$ is removed.

Below we evaluate the conditional probability 
$P(L>\ell | L>\ell-1)$,
namely the probability that the shortest path length between a pair of random
nodes $i$ and $j$ is larger than $\ell$, given that it is larger than $\ell-1$,
using the RPA
\cite{Katzav2015,Nitzan2016,Bonneau2017}.
Since each node in the network (e.g. node $i$) has $c$ neighbors, 
the boundary condition at $\ell=1$ is given by

\begin{equation}
P(L>1|L>0) = 1 - \frac{c}{N-1}.
\end{equation}

\noindent
For $\ell \ge 2$ the conditional probability
can be expressed in the form

\begin{equation}
P(L>\ell | L>\ell-1) = \widetilde P(L>\ell-1 | L>\ell-2)^c,
\label{eq:Pell1}
\end{equation}

\noindent
where
$\widetilde P(L>\ell-1 | L>\ell-2)$
is the conditional distribution of shortest path lengths between 
a node $\tilde i'$ selected via a random edge $e$ and a random node $j$
in the reduced network in which the edge $e$ is removed.
The conditional distribution
$\widetilde P(L>\ell-1 | L>\ell-2)$
can be further expressed in the form

\begin{equation}
\widetilde P(L>\ell-1 | L>\ell-2) = \widetilde P(L>\ell-2 | L>\ell-3)^{c-1},
\label{eq:Pell2}
\end{equation}

\noindent
where the power of $c-1$ reflects the fact that in the reduced network
the node $\tilde i'$ is of degree $c-1$.
In general, the conditional distribution
$\widetilde P(L>\ell-m | L>\ell-m-1)$
is given by

\begin{equation}
\widetilde P(L>\ell-m | L>\ell-m-1) = \widetilde P(L>\ell-m-1 | L>\ell-m-2)^{c-1},
\label{eq:Pell3}
\end{equation}

\noindent
where $m=1,2,\dots,\ell-2$.
This cascade of recursion equations eventually leads to
$\widetilde P(L>1 | L>0)$,
which can be evaluated directly and is given by

\begin{equation}
\widetilde P(L>1 | L>0) =   1 - \frac{c-1}{N-1}.
\label{eq:Pell4}
\end{equation}

\noindent
For sufficiently large $N$ and for $c/N \ll 1$,
Eq. (\ref{eq:Pell4}) can be replaced by

\begin{equation}
\widetilde P(L>1 | L>0) = \exp \left( - \frac{c-1}{N-1} \right).
\label{eq:Pell5}
\end{equation}

\noindent
Inserting 
$\widetilde P(L>\ell-1 | L>\ell-2)$
from Eq. (\ref{eq:Pell2}) into Eq. (\ref{eq:Pell1}),
we obtain

\begin{equation}
P(L>\ell | L>\ell-1) = \widetilde P(L>\ell-2 | L>\ell-3)^{c(c-1)}.
\label{eq:Pell6}
\end{equation}

\noindent
By repeatedly inserting 
$\widetilde P(L>\ell-m | L>\ell-m-1)$
from Eq. (\ref{eq:Pell3}) into the right hand side of Eq. (\ref{eq:Pell6}),
we obtain

\begin{equation}
P(L>\ell | L>\ell-1) = \widetilde P(L>\ell-m-1 | L>\ell-m-2)^{c(c-1)^{m}}.
\label{eq:Pell7}
\end{equation}

\noindent
Finally, for $m=\ell-2$ we obtain

\begin{equation}
P(L>\ell | L>\ell-1) = \exp \left[ - \frac{c(c-1)^{\ell-1}}{N-1} \right].
\label{eq:Pell8}
\end{equation}

\noindent
Similarly, the conditional distribution $\widetilde P(L>\ell | L>\ell-1)$
is given by

\begin{equation}
\widetilde P(L>\ell | L>\ell-1) = \exp \left[ - \frac{(c-1)^{\ell}}{N-1} \right].
\label{eq:Pell8t}
\end{equation}

The tail distribution $P(L>\ell)$ is obtained from

\begin{equation}
P(L>\ell) = \prod_{\ell'=1}^{\ell} P(L>\ell' | L>\ell'-1).
\label{eq:PellTail1}
\end{equation}

\noindent
Inserting $P(L>\ell' | L>\ell'-1)$ from Eq. (\ref{eq:Pell8}) into Eq. (\ref{eq:PellTail1})
and replacing $N-1$ by $N$,
we obtain

\begin{equation}
P(L>\ell) = 
\exp \left[ - \frac{c}{N} \sum_{\ell'=1}^{\ell} (c-1)^{\ell'-1} \right].
\end{equation}

\noindent
Carrying out the summation, we obtain

\begin{equation}
P(L> \ell) = 
\left\{
\begin{array}{ll}
\exp \left[ - \eta \left( e^{b \ell} - 1 \right) \right]  & \ \ \  \ell \ge 0 \\
1 & \ \ \ \ell < 0,
\end{array}
\right.  
\label{eq:taildist}
\end{equation}

\noindent
where 

\begin{equation}
\eta=\frac{c}{(c-2)N},
\label{eq:eta}
\end{equation}

\noindent
and 

\begin{equation}
b=\ln(c-1).
\label{eq:b}
\end{equation}

\noindent
The tail distribution $P(L>\ell)$, given by Eq. (\ref{eq:taildist}),
is a discrete version of the Gompertz distribution
 \cite{Gompertz1825,Shklovskii2005}.
It is in agreement with the DSPL of RRGs obtained in Ref.
\cite{Hofstad2005}.

\begin{figure}
\centerline{
\includegraphics[width=7.5cm]{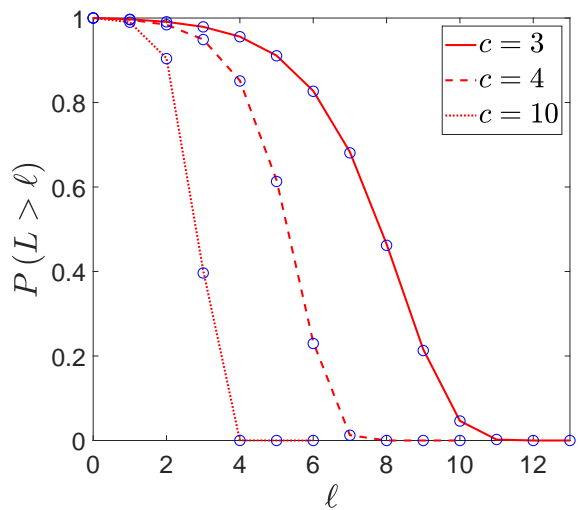}
}
\caption{
Analytical results  
for the tail distribution  
of shortest path lengths
$P(L>\ell)$ for ensembles of RRGs of size $N=1000$ and degrees $c=3$ (solid line),
$c=4$ (dashed line) and $c=10$ (dotted line),
obtained from Eqs. (\ref{eq:taildist})-(\ref{eq:b}).
The analytical results are in very good agreement with the results obtained 
from computer simulations (circles).
As $c$ is increased, the sigmoid function shifts to the left,
which implies that distances in the network become shorter.
It also becomes steeper, which implies that the  
DSPL becomes narrower.
}
\label{fig:1}
\end{figure}

In Fig. \ref{fig:1} we present analytical results for the tail distribution $P(L>\ell)$
of shortest path lengths
for ensembles of RRGs of size $N=1000$ and degrees $c=3$ (solid line),
$c=4$ (dashed line) and $c=10$ (dotted line).
The analytical results,
obtained from Eqs. (\ref{eq:taildist})-(\ref{eq:b}),
are in very good agreement with the results obtained 
from computer simulations (circles).
The tail distribution, which follows a discrete Gompertz distribution,
exhibits a monotonically decreasing sigmoid-like shape,
or a smoothed Heaviside step function.
As $c$ is increased, the sigmoid function shifts to the left,
which implies that distances in the network become shorter.
The sigmoid function also becomes steeper, which implies that the  
DSPL becomes narrower. This means that for sufficiently large
values of $c$ the majority of pairs of nodes are at equal distance from each other.

The probability mass function of the DSPL  
can be recovered from the tail distribution by

\begin{equation}
P(L=\ell) = P(L > \ell - 1) - P(L > \ell).
\label{eq:PL}
\end{equation}

\noindent
As mentioned in the introduction, 
it reflects the shell structure around a random node $i$,
where the $\ell$th shell consists of the nodes that reside
at a distance $\ell$ from $i$.
The expected number of nodes in the $\ell$th shell is
given by

\begin{equation}
n(\ell) = (N-1) P(L=\ell).
\end{equation}

\noindent
The Gompertz distribution frequently appears in the analysis of
life spans and survival. It describes situations in which the mortality
rate grows exponentially with time.
The shell structure around a random node can be interpreted in
terms of survival probabilities. 
To explain this property we express Eq. (\ref{eq:PellTail1}) in the
form

\begin{equation}
P(L>\ell) = P(L>\ell | L>\ell-1) P(L>\ell-1).
\end{equation}

\noindent
This implies that the probability of a random node to remain outside the first 
$\ell$  shells is given by the probability to be outside the first
$\ell-1$ shells times a survival probability, which is given by Eq. (\ref{eq:Pell8}).
The complementary probability $P(L=\ell | L>\ell-1) = 1 - P(L>\ell | L>\ell-1)$
is the probability that such random node ends up in the $\ell$th shell.
In the context of survival analysis this probability corresponds to the 
hazard function or the mortality rate.

\section{The moment generating function of the DSPL}

In this section we apply a systematic method for the calculation of  
moments of the DSPL, which is based on the discrete Laplace
transform of the tail distribution $P(L>\ell)$.
The discrete Laplace transform of some function $f(\ell)$ is given by

\begin{equation}
\mathcal{L} \{ f \}(s) = \sum_{\ell=0}^{\infty}
e^{-s \ell} f(\ell),
\end{equation}

\noindent
which is related to the one-sided Z-transform and to the starred transform
\cite{Phillips2015}.
Inserting $f(\ell)=P(L>\ell)$ we obtain

\begin{equation}
\mathcal{L} \{ P(L>\ell) \}(s) = \sum_{\ell=0}^{N-2}
e^{-s \ell} P(L>\ell),
\label{eq:LapPell1}
\end{equation}

\noindent
where the summation is truncated above $N-2$.
This reflects the fact that in a network that consists of $N$ nodes, $P(L>\ell)=0$ for $\ell \ge N-1$.
In spite of this fact, note that Eq. (\ref{eq:taildist}) yields non-zero results for 
the probabilities $P(L>\ell)$ even for $\ell \ge N-1$.
However, these probabilities are vanishingly small.
Therefore, the upper limit of the sum in Eq. (\ref{eq:LapPell1}) can be
changed from $N-2$ to $\infty$ without affecting the results.
We also note that the tail distribution $P(L>\ell)$ is 
defined as the distribution of distances between pairs
of distinct nodes. Thus, it satisfies $P(L>0)=1$.
Using these observations, Eq. (\ref{eq:LapPell1})
can be written in the form

\begin{equation}
\mathcal{L} \{ P(L>\ell) \}(s) = 1 + \sum_{\ell=1}^{\infty}
e^{-s \ell} P(L>\ell).
\label{eq:LapPell2a}
\end{equation}

\noindent
The sum on the right hand side of Eq. (\ref{eq:LapPell2a}) can be approximated by 
the corresponding integral. However, a more accurate result can be obtained using
the Euler-Maclaurin formula, in which the difference between the sum and the integral
is systematically approximated in terms of derivatives of the integrand, which are evaluated
at the end-points of the interval
\cite{Apostol1999,Bender1999}.
Applying the Euler-Maclaurin formula
(equation 2.10.1 in Ref. \cite{Olver2010}),
we obtain 

\begin{eqnarray}
\mathcal{L} \{ P(L>\ell) \}(s) &=&
1 + \int_{0}^{\infty} e^{-s \ell} P(L>\ell) d \ell
- \frac{1}{2} 
\nonumber \\
&-& \sum_{k=1}^{\infty} \frac{ B_{2k} }{(2k)!}
   \frac{d^{2k-1}}{d \ell^{2k-1}}   \left[ e^{-s \ell} P(L>\ell) \right] \bigg\vert_{\ell=0},
\label{eq:LaplaceEM}
\end{eqnarray}

\noindent
where $B_{2k}$ is the Bernoulli number of order $2k$ 
\cite{Olver2010}.
Carrying out the integration, we obtain

\begin{equation}
\int_{0}^{\infty} e^{-s \ell} P(L>\ell) d \ell =
\frac{ \exp \left[ \left( \frac{c}{c-2} \right) \frac{1}{N} \right] }{\ln (c-1) }
{\rm E}_{1 + \frac{s}{\ln (c-1)}} \left[ \left( \frac{c}{c-2} \right) \frac{1}{N} \right],
\end{equation}

\noindent
where ${\rm E}_n(x)$ is the generalized exponential integral
(Eq. 8.19.3 in Ref. \cite{Olver2010}).
In order to evaluate the sum in Eq. (\ref{eq:LaplaceEM}), 
we expand the summand
up to first order in $1/N$.
Inserting $P(L>\ell)$ from Eq. (\ref{eq:taildist}), we obtain

\begin{equation}
e^{-s \ell} P(L>\ell) =
e^{-s \ell} - \left( \frac{c}{c-2} \right) \frac{1}{N} \left[ e^{-(s-b)\ell} - e^{-s \ell} \right] 
+ \mathcal{O} \left( \frac{1}{N^2} \right),
\label{eq:Lapterm}
\end{equation}

\noindent
where $b=\ln(c-1)$.
Expanding Eq. (\ref{eq:Lapterm}) in powers of $\ell$,
we obtain

\begin{eqnarray}
e^{-s \ell} P(L>\ell) &=&
1 + \left[ 1 + \left( \frac{c}{c-2} \right) \frac{1}{N} \right]
\sum_{r=1}^{\infty} \frac{(-s)^r}{r!} \ell^r 
\nonumber \\
&-& \left( \frac{c}{c-2} \right) \frac{1}{N} \sum_{r=1}^{\infty} \frac{(b-s)^r}{r!} \ell^r 
+ \mathcal{O} \left( \frac{1}{N^2} \right).
\label{eq:Lapterm2}
\end{eqnarray}

\noindent
Differentiating Eq. (\ref{eq:Lapterm2}) $(2k-1)$ times    with respect to $\ell$ 
and evaluating the result at $\ell=0$, we obtain

\begin{eqnarray}
\frac{d^{2k-1}}{d \ell^{2k-1}}   \left[ e^{-s \ell} P(L>\ell) \right] \bigg\vert_{\ell=0} &=&
-\left[ 1 + \left( \frac{c}{c-2} \right) \frac{1}{N} \right] s^{2k-1}
\nonumber \\
&+& \left( \frac{c}{c-2} \right) \frac{1}{N} (s-b)^{2k-1}
+ \mathcal{O} \left( \frac{1}{N^2} \right).
\label{eq:Lapterm3}
\end{eqnarray}

\noindent
Inserting the right hand side of Eq. (\ref{eq:Lapterm3}) into Eq. (\ref{eq:LaplaceEM})
and using the identity
(Eq. 4.19.6 in Ref. \cite{Olver2010}) 

\begin{equation}
\sum_{k=1}^{\infty} 
\frac{ B_{2k} x^{2k-1} }{(2k)!} =
\frac{1}{2} \coth \left( \frac{x}{2} \right) - \frac{1}{x},
\end{equation}

\noindent
we obtain

\begin{eqnarray}
\mathcal{L} \{ P(L>\ell) \}(s) &=&
\frac{ \exp \left[ \left( \frac{c}{c-2} \right) \frac{1}{N} \right] }{\ln (c-1) }
{\rm E}_{1 + \frac{s}{\ln (c-1)}} \left[ \left( \frac{c}{c-2} \right) \frac{1}{N} \right]
+ \frac{1}{2} 
\nonumber \\
&+&
\left[ \frac{1}{2} \coth \left( \frac{s}{2} \right) - \frac{1}{s}  \right]
\left[ 1 + \left( \frac{c}{c-2} \right) \frac{1}{N} \right]
\nonumber \\
&-& \left( \frac{c}{c-2} \right) \frac{1}{N} \left[ \frac{1}{2} \coth \left( \frac{s-b}{2} \right) - \frac{1}{s-b}  \right]
\nonumber \\
&+& \mathcal{O} \left( \frac{1}{N^2} \right).
\label{eq:LapPell2}
\end{eqnarray}

\noindent
In the large network limit Eq. (\ref{eq:LapPell2}) can be reduced to

\begin{eqnarray}
\mathcal{L} \{ P(L>\ell) \}(s) &=&
\frac{1}{ \ln (c-1) } {\rm E}_{1 + \frac{s}{\ln (c-1)}}
\left[ \left( \frac{c}{c-2} \right) \frac{1}{N} \right]
\nonumber \\
&+& \frac{1}{2} 
+
\left[ \frac{1}{2} \coth \left( \frac{s}{2} \right) - \frac{1}{s}  \right]
+ {\mathcal O} \left( \frac{1}{N} \right).
\end{eqnarray}

The moment generating function of the DSPL is denoted by

\begin{equation}
M(s) = \sum_{\ell=0}^{\infty} e^{s \ell} P(L=\ell).
\label{eq:MGF1}
\end{equation}

\noindent
Inserting $P(L=\ell)$ from Eq. (\ref{eq:PL}) into Eq. (\ref{eq:MGF1}),
we can rewrite $M(s)$ in terms of the tail distribution in the form

\begin{equation}
M(s) = 1 + (e^s-1) \sum_{\ell=0}^{\infty} e^{s \ell} P(L>\ell).
\end{equation}

\noindent
Using Eq. (\ref{eq:LapPell1}) we express the moment generating function in
terms of the Laplace transform, namely

\begin{equation}
M(s) = 1 + (e^s-1)  \mathcal{L} \{ P(L>\ell) \}(-s).
\label{eq:MsLPl}
\end{equation}

\noindent
In the sections below we use the moment generating
function to obtain closed form expressions for the mean and 
variance of the DSPL.

\section{The mean distance}

The mean distance $\langle L \rangle$
between pairs of distinct nodes in an RRG
can be calculated using the tail-sum formula 
\cite{Pitman1993}

\begin{equation}
\langle L \rangle = \sum_{\ell=0}^{N-2} P(L > \ell),
\label{eq:<L>}
\end{equation}

\noindent
where $P(L>\ell)$ is given by Eq. (\ref{eq:taildist}).
Since Eq. (\ref{eq:taildist}) is highly accurate for $c \ll N$, Eq. (\ref{eq:<L>}) is expected
to yield accurate results for the mean distance except for the limit of very dense networks.
However, this expression provides little insight on the behavior
of $\langle L \rangle$ and its dependence on $c$ and $N$.
Since $P(L>\ell)$, given by Eq. (\ref{eq:taildist}), is vanishingly small for 
$\ell \ge N-1$, Eq. (\ref{eq:<L>}) can be replaced by 

\begin{equation}
\langle L \rangle = \sum_{\ell=0}^{\infty} P(L > \ell).
\label{eq:<L>inf}
\end{equation}

\noindent
Below we use the moment generating function $M(s)$ to 
obtain a closed-form expression for $\langle L \rangle$,
which is valid for sufficiently large networks.
It is given by

\begin{equation}
\langle L \rangle = \frac{d M(s)}{d s} \Bigg\vert_{s=0}.
\label{eq:LMs}
\end{equation}

\noindent
Inserting $M(s)$ from Eqs. (\ref{eq:MsLPl})
and (\ref{eq:LapPell2}) into Eq. (\ref{eq:LMs})
and carrying out the differentiation, we obtain
   
\begin{eqnarray}
\langle L \rangle &=&   
\frac{ \exp \left[ \left( \frac{c}{c-2} \right) \frac{1}{N} \right] }{\ln (c-1)}
{\rm E_1} \left[ \left( \frac{c}{c-2} \right) \frac{1}{N} \right]
+ \frac{1}{2}
\nonumber \\
&+&
\frac{1}{2}
\left( \frac{c}{c-2} \right) 
\frac{1}{N} 
\left\{ \coth \left[ \frac{\ln (c-1) }{2} \right] - \frac{2}{\ln (c-1)} \right\}
+ \mathcal{O} \left(  \frac{1}{N^2}  \right),
\label{eq:EM5}
\end{eqnarray}

\noindent
where ${\rm E_1}(x)$ is the exponential integral,
also denoted as ${\rm Ei}(x)$
\cite{Olver2010}.
Expanding the right hand side of Eq. (\ref{eq:EM5}) in 
powers of $1/N$,
we obtain

\begin{eqnarray}
\langle L \rangle &=&   
\frac{\ln N}{\ln (c-1)} 
+ \frac{1}{2}
- \frac{ \ln \left( \frac{c}{c-2} \right) + \gamma}{\ln (c-1)}  
+
\frac{1}{\ln (c-1)}
\left( \frac{ c }{c-2} \right)  \frac{\ln N}{N} \ \ \ \ \ \ \ 
\nonumber \\
&+& \frac{c}{c-2} \left\{ \frac{ \ln \left( \frac{ c-2 }{ c } \right) - \gamma}{\ln(c-1)}
+ \frac{1}{2} \coth \left[ \frac{\ln (c-1)}{2} \right] \right\} \frac{1}{N}
+ \mathcal{O} \left(  \frac{\ln N}{N^2}  \right),  
\label{eq:EM6}
\end{eqnarray}

\noindent
where $\gamma = 0.577...$ is the Euler-Mascheroni constant
\cite{Olver2010,Finch2003}.
For sufficiently large values of $N$, the corrections of orders $\ln N/N$ and $1/N$
can be neglected, leading to

\begin{equation}
\langle L \rangle  =    
\frac{\ln N}{\ln (c-1)} 
+ \frac{1}{2}
- \frac{ \ln \left( \frac{c}{c-2} \right) + \gamma}{\ln (c-1)}  
+ \mathcal{O} \left(  \frac{\ln N}{N}  \right).
\label{eq:EM7}
\end{equation}

\begin{figure}
\centerline{
\includegraphics[width=7.5cm]{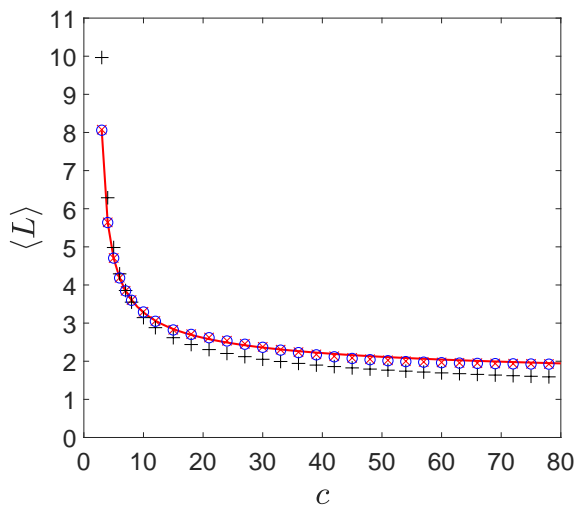}
}
\caption{
Analytical results (solid line) for the mean distance $\langle L \rangle$
between pairs of random nodes in 
RRGs of size $N=1000$ as a function of the degree $c$,
obtained from Eq. (\ref{eq:EM7}).
The results obtained from
direct numerical evaluation of the sum in Eq. (\ref{eq:<L>inf})
are also shown ($\times$ symbols),
where $P(L>\ell)$ is taken from Eq. (\ref{eq:taildist}).
The direct evaluation results
are found to be in very good agreement with the results
obtained from computer simulations (circles).
The analytical results 
are in very good agreement with the results obtained from
direct evaluation and computer simulations,
except for small oscillatory discrepancies discussed in the text.
The widely known result of $\langle L \rangle = \ln N / \ln(c-1)$
is also shown ($+$ symbols), indicating that the subleading term
presented in Eq. (\ref{eq:EM7}) is important.
The subleading correction is found to be negative for $c \le 7$ and positive
for $c>7$.
}
\label{fig:2}
\end{figure}

In Fig. \ref{fig:2} we present 
analytical results (solid line) for the mean distance $\langle L \rangle$ 
between pairs of nodes in an
RRG of size $N=1000$ as a function of the degree $c$,
obtained from the closed-form (CF) expression of Eq. (\ref{eq:EM7}).
We also present the results ($\times$ symbols) obtained from
a direct numerical evaluation (DE) of the sum in Eq. (\ref{eq:<L>inf}),
where $P(L>\ell)$ is taken from Eq. (\ref{eq:taildist}).
Since the discrete Gompertz distribution provides highly accurate results for the DSPL,
the results obtained from
the direct numerical evaluation of Eq. (\ref{eq:<L>inf}) are expected to be accurate.
Indeed, they are found to be in very good agreement with the results
obtained from computer simulations (circles).
The analytical results obtained from Eq. (\ref{eq:EM7})
are in very good agreement with the results obtained from
direct evaluation and computer simulations.
The widely known result of $\langle L \rangle = \ln N / \ln(c-1)$,
given by Eq. (\ref{eq:Leading}),
is also shown ($+$ symbols), indicating that the subleading correction 
presented in Eq. (\ref{eq:EM7}) is important.
However, close inspection of Fig. \ref{fig:2} reveals that
on top of the overall trend, the mean distance $\langle L \rangle$ 
exhibits an oscillatory behavior as a function of $c$,
which is not captured by Eq. (\ref{eq:EM7}).
The amplitude of these oscillations is negligible in the sparse-network limit and 
becomes detectable for dense networks.
Moreover, the period of the oscillations increases as $c$ is increased.

Interestingly, in the dense-network limit the 
direct numerical evaluation of the
mean distance $\langle L \rangle$
can be done using an approximate form of the tail-sum formula of Eq. (\ref{eq:<L>inf}).
It is given by

\begin{equation}
\langle L \rangle \simeq (r-1) + P(L>r-1) + P(L>r) + P(L>r+1),
\label{eq:Ldense}
\end{equation}
 
\noindent
where

\begin{equation}
r = \bigg\lfloor \frac{\ln N}{\ln(c-1)} \bigg\rfloor,
\label{eq:x}
\end{equation}

\noindent
and $\lfloor x \rfloor$ is the integer part of $x$.
The first term on the right hand side of Eq. (\ref{eq:Ldense})
accounts for the sum over the probabilities $P(L>\ell)$ for
$\ell=0,1,\dots,r-2$, which in the dense-network limit can be
approximated by $1$. The next three terms account for the
range of distances in which the tail distribution decreases 
sharply. In the dense-network limit this range is narrow, 
while the probabilities $P(L>\ell)$ for $\ell \ge r+2$
are negligible.

To analyze the oscillations, it is convenient to consider the difference
$\langle L \rangle_{\rm DE} - \langle L \rangle_{\rm CF}$,
where $\langle L \rangle_{\rm DE}$ is the mean distance
obtained from direct numerical evaluation of the mean of the discrete
Gompertz distribution by Eq. (\ref{eq:<L>inf}),
while $\langle L \rangle_{\rm CF}$ is the mean distance obtained
from the closed-form expression of Eq. (\ref{eq:EM7}).

\begin{figure}
\centerline{
\includegraphics[width=7.5cm]{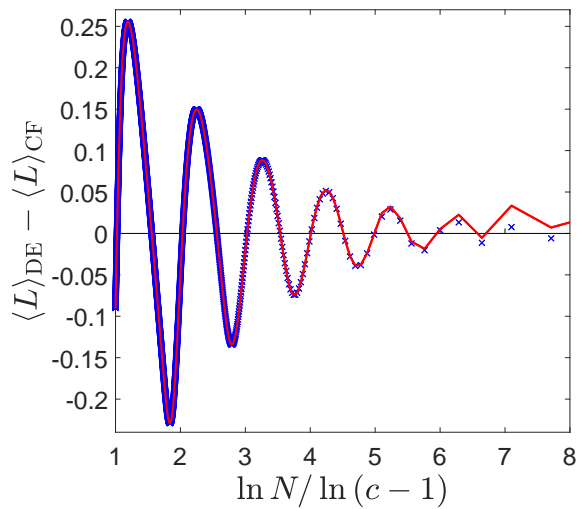}
}
\caption{
The difference 
$\langle L \rangle_{\rm DE} - \langle L \rangle_{\rm CF}$ 
($\times$ symbols)
as a function of
$\ln N/\ln(c-1)$ where the network size is fixed at $N=10^6$ and
the mean degree $c$ is varied.
The mean distance $\langle L \rangle_{\rm DE}$ is obtained from Eq. (\ref{eq:<L>inf})
and $\langle L \rangle_{\rm CF}$ is obtained from Eq. (\ref{eq:EM7}).
This difference exhibits oscillations
as a function of $\ln N/\ln(c-1)$, whose wavelength is equal to $1$.
The amplitude of the oscillations decreases as
$\ln N/\ln(c-1)$ is increased. 
This implies that the oscillations are negligible in the sparse-network limit
and become detectable as the network becomes more dense.
We also present approximated results (solid line) in which $\langle L \rangle_{\rm DE}$
is evaluated using Eq. (\ref{eq:Ldense}).
The two curves are found to be in very good agreement except for the
limit of sparse networks in which Eq. (\ref{eq:Ldense}) is not expected to apply.
In general, the amplitude of the oscillations is very small compared to the mean distance $\langle L \rangle$.
Thus, Eq. (\ref{eq:EM7}) provides a very good approximation for $\langle L \rangle$.
}
\label{fig:3}
\end{figure}

In Fig. \ref{fig:3} we present ($\times$ symbols) the difference 
$\langle L \rangle_{\rm DE} - \langle L \rangle_{\rm CF}$ 
as a function of
$\ln N/\ln(c-1)$ where the network size $N$ is fixed at 
% (a) $N=10^3$; and (b) 
$N=10^6$ and the mean degree $c$ is varied.
The mean distance
$\langle L \rangle_{\rm DE}$ is obtained from Eq. (\ref{eq:<L>inf})
and $\langle L \rangle_{\rm CF}$ is obtained from Eq. (\ref{eq:EM7}).
It is found that  this difference  exhibits oscillations
as a function of $\ln N/\ln(c-1)$, whose wavelength is equal to $1$.
The amplitude of the oscillations decreases as
$\ln N/\ln(c-1)$ is increased. 
The difference 
$\langle L \rangle_{\rm DE} - \langle L \rangle_{\rm CF}$
vanishes at integer and half-integer values of $\ln N/\ln(c-1)$.
It is found that the maxima of the oscillations take place when the fractional part  
of $\ln N/\ln(c-1)$ is approximately $1/4$, while the minima take place 
when the fractional part is approximately $3/4$.
We also present approximated results (solid line) in which $\langle L \rangle_{\rm DE}$
is evaluated using Eq. (\ref{eq:Ldense}).
The two curves are found to be in very good agreement except for the
limit of sparse networks in which Eq. (\ref{eq:Ldense}) is not expected to 
provide accurate results.
It is important to note that
the amplitude of the oscillations is very small compared to the mean distance $\langle L \rangle$.
Thus, Eq. (\ref{eq:EM7}) provides a very good approximation for $\langle L \rangle$.

\begin{figure}
\centerline{
\includegraphics[width=7.5cm]{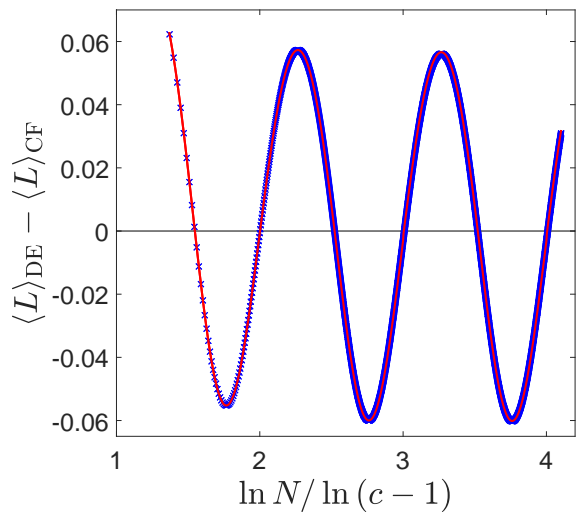}
}
\caption{
The difference 
$\langle L \rangle_{\rm DE} - \langle L \rangle_{\rm CF}$ 
($\times$ symbols)
as a function of
$\ln N/\ln(c-1)$ where the mean degree $c$ is fixed at $c=30$ and the
network size $N$ is varied.
The mean distance $\langle L \rangle_{\rm DE}$ is obtained from Eq. (\ref{eq:<L>inf})
and $\langle L \rangle_{\rm CF}$ is obtained from Eq. (\ref{eq:EM7}).
This difference exhibits oscillations
as a function of $\ln N/\ln(c-1)$, whose wavelength is equal to $1$ 
and the amplitude is a constant.
We also present approximated results (solid line) in which $\langle L \rangle_{\rm DE}$
is evaluated using Eq. (\ref{eq:Ldense}).
The two curves are found to be in very good agreement except for the
limit of sparse networks in which Eq. (\ref{eq:Ldense}) is not expected to apply.
In general, the amplitude of the oscillations is very small compared to the mean distance $\langle L \rangle$.
Thus, Eq. (\ref{eq:EM7}) provides a very good approximation for $\langle L \rangle$.
}
\label{fig:4}
\end{figure}

In Fig. \ref{fig:4} we present ($\times$ symbols) the difference 
$\langle L \rangle_{\rm DE} - \langle L \rangle_{\rm CF}$ 
as a function of
$\ln N/\ln(c-1)$, where the degree $c$ is fixed at 
$c=30$ and the network size $N$ is varied.
% (a) $N=10^3$; and (b) $N=10^6$,
The mean distance
$\langle L \rangle_{\rm DE}$ is obtained from Eq. (\ref{eq:<L>inf})
and $\langle L \rangle_{\rm CF}$ is obtained from Eq. (\ref{eq:EM7}).
It is found that  this difference  exhibits oscillations
as a function of $\ln N/\ln(c-1)$, whose wavelength is equal to $1$.
For sufficiently large values of $N$ the amplitude of the oscillations is a constant.
The difference 
$\langle L \rangle_{\rm DE} - \langle L \rangle_{\rm CF}$
vanishes at integer and half-integer values of $\ln N/\ln(c-1)$.
It is found that the maxima of the oscillations take place when the fractional part  
of $\ln N/\ln(c-1)$ is approximately $1/4$, while the minima take place 
when the fractional part is approximately $3/4$.
We also present approximated results (solid line) in which $\langle L \rangle_{\rm DE}$
is evaluated using Eq. (\ref{eq:Ldense}).
The two curves are found to be in very good agreement.
% except for the
%limit of sparse networks in which Eq. (\ref{eq:Ldense}) is not expected to 
%provide accurate results.

The oscillatory behavior presented in Fig. \ref{fig:4} 
and specifically the fact that the wavelength is $1$ and the amplitude is constant,
implies that the difference
$\langle L \rangle_{\rm DE} - \langle L \rangle_{\rm CF}$ 
depends on $N$ only via the fractional part

\begin{equation}
\phi = \frac{ \ln N }{ \ln (c-1) } -  \bigg\lfloor \frac{ \ln N }{ \ln (c-1) } \bigg\rfloor,
\label{eq:phi}
\end{equation}

\noindent
which can be considered as a phase and takes values in the range $0 \le \phi < 1$.
Below we show that for sufficiently large $N$ this can be rigorously justified.
In the large $N$ limit, where $\eta \ll 1$, Eq. (\ref{eq:taildist}) can be
reduced to

\begin{equation}
P(L> \ell) = \exp \left( - \eta   e^{b \ell}    \right),
\label{eq:taildist2}
\end{equation}

\noindent
which can also be written in the form

\begin{equation}
P(L> \ell) = \exp \left( -   e^{b \ell + \ln \eta}    \right).
\label{eq:taildist3}
\end{equation}

\noindent
Using this expression, one can show that

\begin{equation}
P(L>\ell+r) = 
\left\{
\begin{array}{ll}
\exp \left[ - \left( \frac{c}{c-2} \right) e^{ \ln (c-1) (\ell-\phi) } \right]  & \ \ \  \ell \ge 0  \\
1  & \ \ \  \ell < 0.
\end{array}
\right.   
\label{eq:PLphi}
\end{equation}

\noindent
This implies that the dependence of $P(L>\ell+r)$ on the network size $N$
is only via the phase $\phi$.

The mean distance can be expressed by

\begin{equation}
\langle L \rangle = \sum_{\ell=- \infty}^{\infty} \ell P(L=\ell),
\end{equation}

\noindent
which is justified because $P(L=\ell)=0$ for $\ell \le 0$.
Shifting the summation variable by $r$ to the left, we obtain
the equivalent expression

\begin{equation}
\langle L \rangle = \sum_{\ell=- \infty}^{\infty} (\ell+r) P(L=\ell+r).
\end{equation}

\noindent
Summing up separately the two terms on the right hand side, we obtain

\begin{equation}
\langle L \rangle = r + \sum_{\ell=-\infty}^{\infty} \ell P(L=\ell+r).
\end{equation}

\noindent
Expressing the probability $P(L=\ell+r)$ in terms of the tail distribution, 
we obtain

\begin{equation}
\langle L \rangle = r + \sum_{\ell=-\infty}^{\infty} \ell 
\left[ P(L>\ell+r-1) - P(L>\ell+r) \right].
\label{eq:<L>7}
\end{equation}

\noindent
Subtracting $\langle L \rangle_{\rm CF}$,
given by Eq. (\ref{eq:EM7})
from Eq. (\ref{eq:<L>7}),
we obtain

\begin{eqnarray}
\langle L \rangle_{\rm DE} - \langle L \rangle_{\rm CF} &=& 
- \phi - \frac{1}{2} + \frac{ \ln \left( \frac{c}{c-2} \right) + \gamma }{\ln (c-1)}
\nonumber \\
&+& \sum_{\ell=-\infty}^{\infty} \ell 
\left[ P(L>\ell+r-1) - P(L>\ell+r) \right].
\label{eq:<L>8}
\end{eqnarray}

\noindent
Inserting the probabilities $P(L>\ell+r)$ from Eq. (\ref{eq:PLphi})
into Eq. (\ref{eq:<L>8}), one finds that
$\langle L \rangle_{\rm DE} - \langle L \rangle_{\rm CF}$
is only a function of the degree $c$ and the phase $\phi$.
This implies that its dependence on the network size $N$ is only
via the phase $\phi$.
Note that the argument above holds when the network size $N$ is
sufficiently large, such that Eq. (\ref{eq:PLphi}) is a good approximation
for Eq. (\ref{eq:taildist}).

\section{The variance of the DSPL}

Using the tail-sum formula, the second moment of the DSPL
can be expressed in the form
\cite{Pitman1993}

\begin{equation}
\langle L^2 \rangle = \sum_{\ell=0}^{N-2}
(2\ell+1) P(L>\ell).
\label{eq:<L2>}
\end{equation}

\noindent
Since the probability $P(L>\ell)$, given by Eq. (\ref{eq:taildist}),
is vanishingly small for $\ell \ge N-1$, the upper limit of the summation can be
changed from $N-2$ to $\infty$, without any noticeable 
change in the result.
Therefore,

\begin{equation}
\langle L^2 \rangle = 1 + \sum_{\ell=1}^{\infty}
(2\ell+1) P(L>\ell).
\label{eq:<L2b>}
\end{equation}

\noindent
The variance of the DSPL is given by

\begin{equation}
{\rm Var}(L) = \langle L^2 \rangle -   \langle L \rangle^2.
\label{eq:Var1}
\end{equation}

\noindent
Inserting $\langle L \rangle$
from Eq. (\ref{eq:<L>inf})
and inserting $\langle L^2 \rangle$ from Eq. (\ref{eq:<L2b>})
into Eq. (\ref{eq:Var1}) yields highly accurate 
results for the variance of the DSPL.
However, this expression provides little insight on the behavior
of the variance and its dependence on $c$ and $N$.

Below we use the moment generating function $M(s)$ to 
obtain a closed-form expression for $\langle L^2 \rangle$,
which is valid for large networks.
It is given by

\begin{equation}
\langle L^2 \rangle = \frac{d^2 M(s)}{d s^2} \Bigg\vert_{s=0}.
\label{eq:LMs2}
\end{equation}

\noindent
Inserting $M(s)$ from Eqs. (\ref{eq:MsLPl}) and (\ref{eq:LapPell2}) into Eq. (\ref{eq:LMs2})
and carrying out the differentiations, we obtain

\begin{eqnarray}
\langle L^2 \rangle &=& 
\frac{ \exp \left[ \frac{c}{(c-2)N} \right]  }{6 [\ln (c-1) ]^2}
\left\{ 6 \left( \ln \left[ \frac{c}{(c-2)N} \right] \right)^2 
+ 12 \gamma \ln \left[ \frac{c}{(c-2)N} \right]
\right.
\nonumber \\
&+& \pi^2  + 6 \gamma^2 
\left.
- \frac{12 c}{(c-2)N} \ 
_3F_3 \left[ \left.
\begin{array}{c}
1,1,1 \\
2,2,2
\end{array}
\right| - \frac{c}{(c-2)N} 
\right] \right\}
\nonumber \\
&+& \frac{ \exp \left[ \frac{c}{(c-2)N} \right] }{\ln (c-1)}
{\rm E_1} \left[ \left( \frac{c}{ c-2 } \right)  \frac{1}{N} \right]
+ \frac{1}{3}
+ \mathcal{O} \left( \frac{1}{N^2} \right).
\label{eq:L2allN}
\end{eqnarray}

\noindent
where 
${\rm E_1}(x)$ is the exponential integral and
$_3F_3[\ ]$
is the generalized hypergeometric function
\cite{Olver2010}.
Performing an asymptotic expansion for large $N$,
we obtain

\begin{eqnarray}
\langle L^2 \rangle &=& 
\frac{1 }{  [\ln (c-1) ]^2 }
\left( \ln \left[ \frac {(c-2)N}{c} \right] \right)^2 
+  \frac{ \ln (c-1) - 2 \gamma }{  [\ln (c-1) ]^2 }
\ln \left[ \frac{(c-2)N}{c} \right]
\nonumber \\
&+& \frac{ \pi^2  + 6 \gamma^2  - 6 \gamma \ln (c-1) }{ 6  [\ln (c-1) ]^2 }
+ \frac{1}{3}
+ \mathcal{O} \left[ \left( \frac{\ln N}{N} \right)^2 \right].
\label{eq:L2largeN}
\end{eqnarray}

\noindent
To obtain the variance of the DSPL we insert $\langle L^2 \rangle$ from Eq. (\ref{eq:L2allN})
and $\langle L \rangle$ from Eq. (\ref{eq:EM5}) into Eq. (\ref{eq:Var1}).
While this result is expected to be relatively precise the resulting expression is complicated.
A simpler expression for the variance can be obtained by inserting
$\langle L^2 \rangle$ from Eq. (\ref{eq:L2largeN})
and $\langle L \rangle$ from Eq. (\ref{eq:EM7}) into Eq. (\ref{eq:Var1}).
In this case the resulting expression can be simplified to the form

\begin{equation}
{\rm Var}(L) = \frac{ \pi^2 }{6 [ \ln(c-1) ]^2 }  + \frac{1}{12}
+ \mathcal{O} \left( \frac{\ln N}{N} \right).
\label{eq:kappa2}
\end{equation}

\noindent
This result implies that except for the limit of very small networks,
the variance ${\rm Var}(L)$ does not depend on the network size $N$
but only on the degree $c$.
Moreover ${\rm Var}(L)$ is a monotonically decreasing function of $c$,
whose largest value, obtained at $c=3$ is ${\rm Var}(L) \simeq 3.51$.
We thus conclude that the DSPL of RRGs is a narrow distribution, whose
width decreases as $c$ is increased.

\begin{figure}
\centerline{
\includegraphics[width=7.0cm]{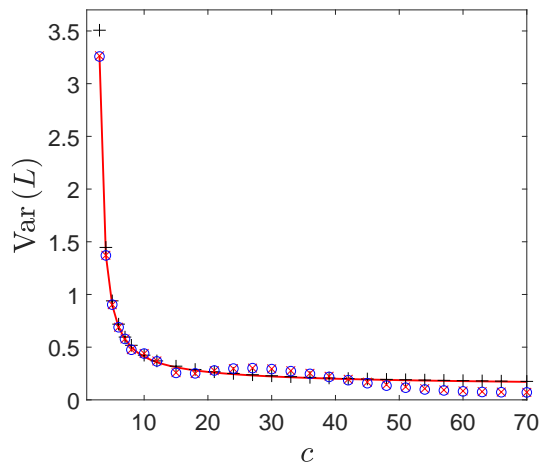}
}
\caption{
Closed form analytical results (solid line) for the variance ${\rm Var}(L)$
of the distribution of shortest path lengths for an RRG of size $N=1000$ as a function of the degree $c$,
obtained from Eq. (\ref{eq:Var1}),
where 
$\langle L \rangle$ is given by Eq. (\ref{eq:EM5})
and 
$\langle L^2 \rangle$ is given by Eq. (\ref{eq:L2allN}).
The results obtained from
direct numerical evaluation of the sums in Eqs. (\ref{eq:<L>inf})
and (\ref{eq:<L2b>})
for  $\langle L \rangle$ and $\langle L^2 \rangle$, respectively
are also shown ($\times$ symbols).
The direct evaluation results are found to be in very good agreement with the results
obtained from computer simulations (circles).
In the regime of sparse networks
the analytical results obtained from 
the closed form expressions of 
Eqs. (\ref{eq:EM5}) and (\ref{eq:L2allN})
are in very good agreement with the results obtained from
direct evaluation and computer simulations.
However, for dense networks
the results obtained from direct evaluation
and computer simulations exhibit oscillations that are
not captured by the closed-form expression.
Instead, the closed-form expression captures the overall trend
of ${\rm Var}(L)$ vs. $c$ as if the oscillations are averaged out.
We also present the results obtained from 
from the simpler expression of Eq. (\ref{eq:kappa2})
($+$ symbols). 
These results are found to be in good agreement with the results obtained 
from the more complete expressions of 
Eqs. (\ref{eq:EM5}) and (\ref{eq:L2allN}),
except for a small discrepancy for very small values of $c$.
}
\label{fig:5}
\end{figure}

In Fig. \ref{fig:5} we present
analytical results (solid line)
for the variance ${\rm Var}(L)$
for RRGs of size $N=1000$ as a function of the degree $c$
obtained from Eq. (\ref{eq:Var1}),
where 
$\langle L \rangle$ is given by Eq. (\ref{eq:EM5})
and 
$\langle L^2 \rangle$ is given by Eq. (\ref{eq:L2allN}).
We also present the results obtained from
a direct numerical evaluation ($\times$ symbols) of the sums in Eqs. (\ref{eq:<L>inf})
and (\ref{eq:<L2b>})
for  $\langle L \rangle$ and $\langle L^2 \rangle$, respectively.
These results are found to be in very good agreement with the results
obtained from computer simulations (circles).
In the regime of sparse networks
the analytical results obtained from 
the closed form expressions of 
Eqs. (\ref{eq:EM5}) and (\ref{eq:L2allN})
are in very good agreement with the results obtained from
direct evaluation and computer simulations.
For dense networks
the results obtained from direct evaluation
and computer simulations exhibit some oscillations that are
not captured by the closed form expressions.
Instead, the closed-form expressions capture the overall trend
of ${\rm Var}(L)$ vs. $c$ as if the oscillations are averaged out.
We also present the results obtained   
from the simpler expression of Eq. (\ref{eq:kappa2})
($+$ symbols). For very small values of $c$,  
Eq. (\ref{eq:kappa2}) is found to slightly 
over-estimate the variance, while for larger values of $c$
it is in very good agreement with the results obtained from
Eqs. (\ref{eq:EM5}) and (\ref{eq:L2allN}).

In the dense-network limit, the tail sum formula (\ref{eq:<L2>})
can be approximated by

\begin{equation}
\langle L^2 \rangle \simeq (r-1)^2 + \sum_{\ell=r-1}^{r+1} (2 \ell + 1) P(L>\ell),
\label{eq:L2dense}
\end{equation}

\noindent
where $r$ is given by Eq. (\ref{eq:x}).
Inserting $\langle L^2 \rangle$ from Eq. (\ref{eq:L2dense})
and $\langle L \rangle$ from Eq. (\ref{eq:Ldense}) into Eq. (\ref{eq:Var1})
and rearranging terms, we obtain

\begin{eqnarray}
{\rm Var}(L) & \simeq 
P(L>r-1) + 3 P(L>r) + 5 P(L>r+1) 
\nonumber \\
&-  \left[ P(L>r-1) + P(L>r) + P(L>r+1) \right]^2.
\label{eq:VarLarge3}
\end{eqnarray}

To analyze the discrepancy between the results obtained from the closed-form
expressions and those obtained from the direct numerical evaluation of the variance,
it is convenient to consider the difference
${\rm Var}_{\rm DE}(L) - {\rm Var}_{\rm CF}(L)$,
where
${\rm Var}_{\rm DE}(L)$
is the variance obtained from the direct numerical evaluation of  
Eqs. (\ref{eq:<L>inf}) and (\ref{eq:<L2b>})
and
${\rm Var}_{\rm CF}(L)$
is the variance obtained from the closed form expressions of Eqs.
(\ref{eq:EM5}) and (\ref{eq:L2allN}).
It turns out that this difference exhibits oscillations
similar to those obtained for $\langle L \rangle_{\rm DE} - \langle L \rangle_{\rm CF}$,
between positive and negative values as a function of $c$.
Moreover, the period and the amplitude of these oscillations
increase as $c$ is increased.

\begin{figure}
\centerline{
\includegraphics[width=7.0cm]{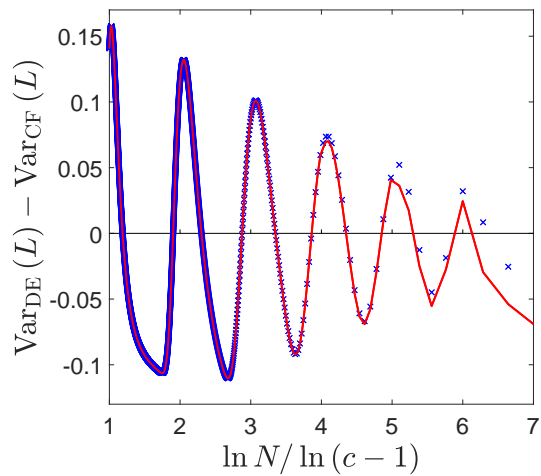}
}
\caption{
The difference  
${\rm Var}_{\rm DE}(L) - {\rm Var}_{\rm CF}(L)$ 
($\times$ symbols)
between the variance obtained from direct numerical evaluation
of Eqs. (\ref{eq:<L>inf}) and (\ref{eq:<L2b>})
and from the closed form expressions given by Eqs. 
(\ref{eq:EM5}) and (\ref{eq:L2allN}),
as a function of
$\ln N/\ln(c-1)$ for $N=10^6$.
This difference exhibits oscillations
as a function of $\ln N/\ln(c-1)$, whose wavelength is equal to $1$.
The amplitude of the oscillations decreases as
$\ln N/\ln(c-1)$ is increased. 
This implies that the oscillations are negligible in the sparse-network limit
and become more pronounced as the network becomes more dense.
We also present approximated results (solid line) 
in which ${\rm Var}_{\rm DE}(L)$ is evaluated using Eq. (\ref{eq:VarLarge3}).
The two curves are found to be in very good agreement except for the
limit of sparse networks in which Eq. (\ref{eq:VarLarge3})
is not expected to provide accurate results.
}
\label{fig:6}
\end{figure}

In Fig. \ref{fig:6} we present ($\times$ symbols) the difference  
${\rm Var}_{\rm DE}(L) - {\rm Var}_{\rm CF}(L)$   as a function of
$\ln N/\ln(c-1)$ for $N=10^6$.
It is found that this difference exhibits oscillations
as a function of $\ln N/\ln(c-1)$, whose wavelength is equal to $1$.
The amplitude of the oscillations decreases as
$\ln N/\ln(c-1)$ is increased. 
It is found that the maxima of the oscillations take place at integer 
values of $\ln N/\ln(c-1)$, while the minima take place at half-integer values.
This means that around integer values of $\ln N/\ln(c-1)$ the closed-form expression 
provides an under-estimated value for ${\rm Var}(L)$, while around half-integer
values of $\ln N/\ln(c-1)$ the closed-form expressions provide an over-estimated 
value for ${\rm Var}(L)$.
We also present approximated results (solid line) 
in which ${\rm Var}_{\rm DE}(L)$ is evaluated using Eq. (\ref{eq:VarLarge3}).
The two curves are found to be in very good agreement except for the
limit of sparse networks in which Eq. (\ref{eq:VarLarge3})
is not expected to provide accurate results.

\begin{figure}
\centerline{
\includegraphics[width=7.0cm]{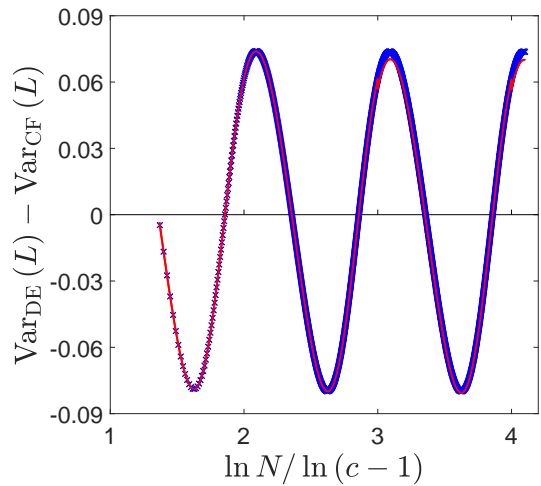}
}
\caption{
The difference  
${\rm Var}_{\rm DE}(L) - {\rm Var}_{\rm CF}(L)$ 
($\times$ symbols)
between the variance obtained from direct numerical evaluation
of Eqs. (\ref{eq:<L>inf}) and (\ref{eq:<L2b>})
and from the closed form expressions given by Eqs. 
(\ref{eq:EM5}) and (\ref{eq:L2allN}),
as a function of
$\ln N/\ln(c-1)$ where the degree is fixed at $c=30$ 
and the network size $N$ is varied.
This difference exhibits oscillations
as a function of $\ln N/\ln(c-1)$, whose wavelength is equal to $1$
and the amplitude is a constant.
We also present approximated results (solid line) 
in which ${\rm Var}_{\rm DE}(L)$ is evaluated using Eq. (\ref{eq:VarLarge3}).
The two curves are found to be in very good agreement except for the
limit of sparse networks in which Eq. (\ref{eq:VarLarge3})
is not expected to provide accurate results.
}
\label{fig:7}
\end{figure}

In Fig. \ref{fig:7} we present ($\times$ symbols) the difference  
${\rm Var}_{\rm DE}(L) - {\rm Var}_{\rm CF}(L)$   as a function of
$\ln N/\ln(c-1)$ where the degree $c$ is fixed at $c=30$ and the 
network size $N$ is varied.
It is found that this difference exhibits oscillations
as a function of $\ln N/\ln(c-1)$, whose wavelength is equal to $1$ 
and the amplitude is a constant.
It is found that the maxima of the oscillations take place at integer 
values of $\ln N/\ln(c-1)$, while the minima take place at half-integer values.
This means that around integer values of $\ln N/\ln(c-1)$ the closed-form expression 
provides an under-estimated value for ${\rm Var}(L)$, while around half-integer
values of $\ln N/\ln(c-1)$ the closed-form expressions provide an over-estimated 
value for ${\rm Var}(L)$.
We also present approximated results (solid line) 
in which ${\rm Var}_{\rm DE}(L)$ is evaluated using Eq. (\ref{eq:VarLarge3}).
The two curves are found to be in very good agreement except for the
limit of sparse networks in which Eq. (\ref{eq:VarLarge3})
is not expected to provide accurate results.
The oscillatory behavior  
presented in Fig. \ref{fig:7} implies that the difference
${\rm Var}_{\rm DE}(L) - {\rm Var}_{\rm CF}(L)$
depends on $N$ only via the phase $\phi$.
This can be justified using an argument similar to the one presented 
for the mean distance at the end of Sec. 5.

\section{Discussion}

In configuration model networks, in the large network limit, the mean distance
can be approximated by Eq. (\ref{eq:<L>conf}).
It is interesting to compare this result to
the corresponding results in models of growing networks.
Here we focus on a model of random networks that grow by node duplication (ND),
introduced in Ref. 
\cite{Lambiotte2016}.
In this model, at each time step a random (mother) node is duplicated.
The daughter node is connected to the mother node and is also connected 
to each one of the neighbors of the mother node with probability $p$.
This model exhibits a phase transition at $p=1/2$ between the  
sparse-network regime at $p<1/2$ and the dense-network regime at $p>1/2$.
In the sparse-network regime,
the mean distance is given by
\cite{Steinbock2017}

\begin{equation}
\langle L_{\rm ND} \rangle = 2 (1-\eta) \ln N + \mathcal{O}(1),
\label{eq:L_ND}
\end{equation}

\noindent
where 
$\eta = p + 2 p^3 + \mathcal{O}(p^4)$.
Thus, the mean distance of a node duplication network
scales logarithmically with $N$ as in the case of configuration
model networks.
However, it was found that for a given network size $N$,
the mean distance $\langle L_{\rm ND} \rangle$
is significantly larger than the mean distance of
a configuration model network with the same degree distribution
\cite{Steinbock2017}.
The variance of the DSPL of ND networks is given by
\cite{Steinbock2017}

\begin{equation}
{\rm Var}(L_{\rm ND}) = 2 (1-\eta) \ln N + \mathcal{O}(1).
\label{eq:V_ND}
\end{equation}

\noindent
Comparing Eqs. (\ref{eq:L_ND}) and (\ref{eq:V_ND}), one observes that
the mean and variance are the same, which is typical to Poisson-like
distributions.
This implies that
the DSPL of ND networks is a broad distribution whose width scales like $(\ln N)^{1/2}$.
This is in contrast to the case of RRGs, in which the variance 
is very small and does not scale with the network size.

In directed ND networks not all the pairs of nodes are connected by directed paths.
Conditioning on pairs of nodes that are connected by directed paths, it was found that
${\mathbb E}[L|L<\infty] \sim \ln N$, which is similar to the result
for undirected ND networks.
However, the variance scales like
${\rm Var}(L) \sim (\ln N)^2$,
which means that the distribution is much broader
\cite{Steinbock2019}.

Another quantity 
which is closely related to the mean distance 
is the mean diameter $\langle D \rangle$ 
\cite{Fernholz2007,Riordan2010,Chung2001}.
Unlike the mean distance $\langle L \rangle$ which is averaged over all
pairs of nodes in each network instance as well as over the ensemble,
the mean diameter is averaged only over the ensemble
(each network instance provides a single value of the diameter).
It was shown that the mean diameter of 
an ensemble of RRGs is given by
\cite{Bollobas1982}

\begin{equation}
\langle D \rangle = \frac{\ln N}{\ln (c-1)} + \frac{\ln \ln N}{\ln (c-1)} +
\mathcal{O}(1).
\end{equation}

\noindent
It thus turns out that the expressions for the
mean distance $\langle L \rangle$ and the mean diameter $\langle D \rangle$
share the same leading term. 
This is not an obvious result.
It appears to reflect the fact that in  the shell structure around a random node, 
most nodes are concentrated in the last few shells.
Such networks, in which most of the distances are almost the same, are called idemetric networks
\cite{Barmpalias2018}.
Having said that, the subleading term for the mean diameter scales like $\ln \ln N$,
unlike the case of $\langle L \rangle$ in which it does not depend on $N$.
This reflects the fact that the diameter of an RRG of size $N$ is 
the maximal distance between all pairs of nodes. 
As $N$ is increased the number of such pairs increases,
making it more probable to find a pair of nodes that are farther away from each other.

\section{Summary}
 
The DSPL of  RRGs follows a discrete Gompertz distribution.
Using the discrete Laplace transform
and the Euler-Maclaurin formula we
derived a closed-form expression for the moment generating function
of the DSPL.
From the moment generating function we obtained expressions for 
the mean distance $\langle L \rangle$ and for the variance ${\rm Var}(L)$ of the DSPL.
More specifically, it was found that the mean distance is given by

\begin{equation}
\langle L \rangle  \simeq    
\frac{\ln N}{\ln (c-1)} 
+ \frac{1}{2}
- \frac{ \ln \left( \frac{c}{c-2} \right) + \gamma}{\ln (c-1)}, 
%+ \mathcal{O} \left(  \frac{\ln N}{N}  \right),
\label{eq:EM7smr}
\end{equation}

\noindent
and the variance of the DSPL is given by

\begin{equation}
{\rm Var}(L) \simeq 
\frac{ \pi^2 }{6 [ \ln(c-1) ]^2 }  + \frac{1}{12}.
%+ \mathcal{O} \left( \frac{1}{N} \right).
\label{eq:kappa2smr}
\end{equation}

\noindent
The result for $\langle L \rangle$ extends known results by adding a
correction term, which yields very good agreement with 
the results obtained from direct numerical evaluation of $\langle L \rangle$
via the tail-sum formula and with the results obtained from computer simulations.
The expression obtained for the variance captures  
the overall dependence of the variance on the degree $c$.
However, it turns out that on top of the overall trend, 
the mean distance $\langle L \rangle$ and
the variance ${\rm Var}(L)$ also exhibit some oscillatory
behavior, which is not captured by the closed-form expressions.
%These oscillations were analyzed and discussed.
The oscillations of the mean and variance 
are due to the discrete nature of the shell structure around a random node.
They reflect the profile of the filling of new shells as $N$ is increased, 
or as $c$ is decreased.
It was shown that these oscillations depend on $N$ only via the phase $\phi$,
defined in Eq. (\ref{eq:phi}).
This implies regular oscillations with wavelength $1$ as a function of $\ln N/\ln (c-1)$,
when $N$ is varied and $c$ is kept fixed.
The results for the mean and variance of the DSPL were compared to the corresponding results 
obtained in other types of random networks.
The relation between the mean distance and the diameter was also discussed.

The authors wish to thank one of the anonymous referees for insightful 
comments and analysis that helped to clarify the nature of the oscillations
of the mean and variance of the DSPL.
This work was supported by the Israel Science Foundation grant no. 
1682/18.

%\newpage
\end{document}